\documentclass[conference]{IEEEtran}
\IEEEoverridecommandlockouts
\usepackage{cite}
\usepackage{amsmath,amssymb,amsfonts}
\usepackage{algorithmic}
\usepackage{graphicx}
\usepackage{textcomp}
\usepackage{xcolor}
\usepackage{balance}
\usepackage[shortcuts,acronym]{glossaries}
\usepackage{booktabs}

\newacronym{mm}{mmWave}{millimeter-wave}
\newacronym{rf}{RF}{radio frequency}
\newacronym{adc}{ADC}{analog-to-digital converter}
\newacronym{cs}{CS}{compressive sensing}
\newacronym{dl}{DL}{deep learning}
\newacronym{sr}{SR}{super-resolution}
\newacronym{ir}{IR}{image-restoration}

\newacronym{gan}{GAN}{generative adversarial network}
\newacronym{cgan}{cGAN}{conditional generative adversarial network}

\newacronym{omp}{OMP}{orthogonal matching pursuit}
\newacronym{amp}{AMP}{approximate message passing }
\newacronym{lamp}{LAMP}{learnable approximate message passing}
\newacronym{ldamp}{LDAMP}{learned
denoising-based approximate message passing}
\newacronym{gmlamp}{GM-LAMP}{Gaussian mixture LAMP}
\newacronym{dnn}{DNN}{deep neural network}
\newacronym{mimo}{MIMO}{multiple input multiple output}
\newacronym{bs}{BS}{base station}
\newacronym{ue}{UE}{user}
\newacronym{ula}{ULA}{uniform linear array}
\newacronym{upa}{UPA}{uniform planar array}
\newacronym{dft}{DFT}{discrete Fourier transform}
\newacronym{eau}{EAU}{enhancement attention unit}

\newacronym{nmse}{NMSE}{normalized mean square error}

\def\BibTeX{{\rm B\kern-.05em{\sc i\kern-.025em b}\kern-.08em
    T\kern-.1667em\lower.7ex\hbox{E}\kern-.125emX}}
\begin{document}

\title{RIDNet Assisted cGAN Based Channel Estimation for One-Bit ADC mmWave MIMO Systems}

\author{\IEEEauthorblockN{Erhan Karakoca\IEEEauthorrefmark{1}\IEEEauthorrefmark{2}\IEEEauthorrefmark{3}, Hasan Nayir\IEEEauthorrefmark{1}\IEEEauthorrefmark{2}\IEEEauthorrefmark{3}, Ali Görçin\IEEEauthorrefmark{3}\IEEEauthorrefmark{4}, Khalid Qaraqe\IEEEauthorrefmark{1}}

\IEEEauthorblockA{\IEEEauthorrefmark{1} Department of Electrical and Computer Engineering, Texas A\&M University at Qatar, Doha, Qatar}

\IEEEauthorblockA{\IEEEauthorrefmark{2} Department of Electronics and Communication Engineering, Istanbul Technical University, {\.{I}}stanbul, Turkey}

\IEEEauthorblockA{\IEEEauthorrefmark{3} Communications and Signal Processing Research (HİSAR) Lab., T{\"{U}}B{\.{I}}TAK B{\.{I}}LGEM, Kocaeli, Turkey}

\IEEEauthorblockA{\IEEEauthorrefmark{4} Department of Electronics and Communication Engineering, Yildiz Technical University, {\.{I}}stanbul, Turkey\\
Emails: \texttt{\{karakoca19, nayir20\}@itu.edu.tr, agorcin@yildiz.edu.tr,}\\
\texttt{{khalid.qaraqe@qatar.tamu.edu}}}}   
\maketitle
\begin{abstract}
The estimation of millimeter-wave (mmWave) massive multiple input multiple output (MIMO) channels becomes compelling when one-bit analog-to-digital converters (ADCs) are utilized.
Furthermore, as the number of antenna increases, pilot overhead scales up to provide consistent channel estimation, eventually degrading spectral efficiency.
This study presents a channel estimation approach that combines a conditional generative adversarial network (cGAN) with a novel blind denoising network with a sparse feature attention mechanism. Performance analysis and simulations show that using a cGAN fused with a feature attention-based denoising neural network significantly enhances the channel estimation performance while requiring less pilot transmission.
\end{abstract}
\begin{IEEEkeywords}
channel estimation, one-bit ADC, massive MIMO, generative adversarial network, feature attention
\end{IEEEkeywords}
\section{Introduction}

The \ac{mm} along with massive \ac{mimo} enables high-capacity communication for next-generation wireless mobile systems.
However, massive \ac{mimo} systems bring additional complexity and cost to the overall system architecture. 
In particular, a large number of \ac{rf} chains and analog-to-digital converters (ADCs) used in massive MIMO systems significantly increase power consumption. 
This difficulty can be handled either by reducing the number of \ac{rf} chains or utilizing one-bit \ac{adc} \cite{li2017channel}.
The significant drawback of one-bit \ac{adc} is that the received signals are highly quantized, making channel estimation a challenging task. 
Therefore, high-accuracy and low-complexity estimation methods using minimal pilot overhead are needed for one-bit ADC massive MIMO systems.

Initial studies for massive MIMO channel estimation have been performed using \ac{cs} methods based on \ac{amp} algorithms which exploit sparsity in \ac{mm} channel \cite{mo2014channel, vila2013expectation}.
In practical scenarios, the performance of the AMP algorithm in channel estimation is bounded by the challenge of determining optimal shrinkage parameters.
As another \ac{cs} method, an expectation-consistent signal recovery algorithm has been proposed to model sparse \ac{mm} channel elements using a Laplacian prior assumption \cite{wang2019channel}, where the proposed technique outperforms AMP-based algorithms.  

The CS methods require higher computational complexity and have lower performance compared to \ac{dl} based approaches which have gained prominence for dealing with a wide range of channel estimation problems.
For instance, \ac{ldamp} \cite{he2018deep}  is proposed for \ac{mm} beamspace channel estimation, where a denoising convolutional neural network (DnCNN) integrated with the AMP algorithm is utilized to reduce channel noise.
Two-stage \ac{dl}-based channel estimation techniques are common in \ac{mm} \ac{mimo} systems when there are limited channel observations \cite{soltani2019deep,chun2019deep,wei2019amp, ji2021mmwave}. The unknown values of the channel matrix are predicted in the first stage, while the noise effects are reduced in the second stage. 
For example, a two-stage \ac{dl} approach is employed to estimate channel state information in \cite{soltani2019deep}, where the unknown values of the channel matrix are estimated from the known pilot symbols using \ac{sr} network while \ac{ir} network is used to minimize the noise effect.  
Furthermore, another two-stage \ac{dl}-based channel estimator is proposed in \cite{chun2019deep} for \ac{mm} \ac{mimo} assuming that the number of antennas is lower than the number of pilots. 
The authors utilize a two-layer neural network and \ac{dnn} for pilot-aided channel estimation, while the second \ac{dnn} is utilized for improving the channel estimation accuracy. 

Recently, \ac{mm} \ac{mimo} channel estimation is considered as a channel generation problem which is solved by using a \ac{gan} in \cite{balevi2021wideband}.
The loss functions of conventional \ac{dnn}s are not well-suited for the one-bit MIMO channel estimation problem. Instead, experimentally tuned generic loss functions (i.e., $\mathcal{L}_1$ or $\mathcal{L}_2$) are utilized for \ac{dnn}s.
On the other hand, \ac{gan}s allow designing a flexible loss function by learning the distribution of features belonging to the data adversarially. 
Most notably, a variation of GAN called conditional \ac{gan} (cGAN) \cite{mirza2014conditional}, allows the model to generate samples conditioned on additional information by its loss function rather than generated randomly as in GANs.
For example, cGAN can be used to generate samples of channel matrices using limited channel observations and pilot symbols, which helps yield more accurate channels. 
c\ac{gan} is applied for one-bit \ac{mm} \ac{mimo} channel estimation in \cite{dong2020channel}, and where the results reveal that c\ac{gan} outperforms even enhanced \ac{cs} methods and provide progress over \ac{dl}-based channel estimation techniques.

Note that the above-mentioned \ac{dnn}-based channel estimation methods have multiple stages. 
Although cGAN can provide slightly more accurate results over \ac{dnn}-based approaches, the channel estimation ability can be further enhanced by employing a second stage. 
Considering that the observations are heavily quantized and reduced, the second stage can further decrease estimation errors across adjacent channel matrix components.
In this paper, we propose a two-stage \ac{dl}-based channel estimation technique inspired by both c\ac{gan} \cite{isola2017image} and real image denoising network (RIDNet)\cite{anwar2019real}, named cGAN-RIDNet.
In the first stage, the coarse channel is reconstructed from highly quantized limited channel observations by having conditional input to cGAN and using two additional regularizations in the loss function.
The second stage further enhances the channel matrix using RIDNet, significantly reducing noise from the coarsely estimated channel matrix.
The numerical results show that RIDNet can successfully learn noise features in both spatial and angular domains. However, the channel estimation accuracy is higher when the RIDNet uses angular domain samples where the channel is represented sparsely. 
The cGAN-RIDNet provides almost identical performance results when the number of pilots is half that of the cGAN.
The computational complexity of the RIDNet can be negligible compared to cGAN since its computational time is almost 1\% of cGAN.


The remainder of this paper is structured as follows: Section II outlines the system model for a one-bit ADC mmWave MIMO channel, while Section III introduces the proposed cGAN-RIDNet architecture. Finally, Section IV presents simulation results, and Section V concludes the paper. Throughout the manuscript, we denote vectors using lowercase bold letters (e.g., ${\mathbf{x}}$), and matrices using uppercase bold letters (e.g., $\mathbf{X}$). The probability density function of the uniform distribution is denoted by $\mathcal{U}$. Additionally, $\mathcal{F}_{\operatorname{x}}(\cdot)$ represents a function of the given subscript x.
The superscript ${*}$ in text and figures denotes the sparse representation of the channel.

\section{The Channel and System Model}

\begin{figure}[t!]
    \centering
    \includegraphics[width=\linewidth]{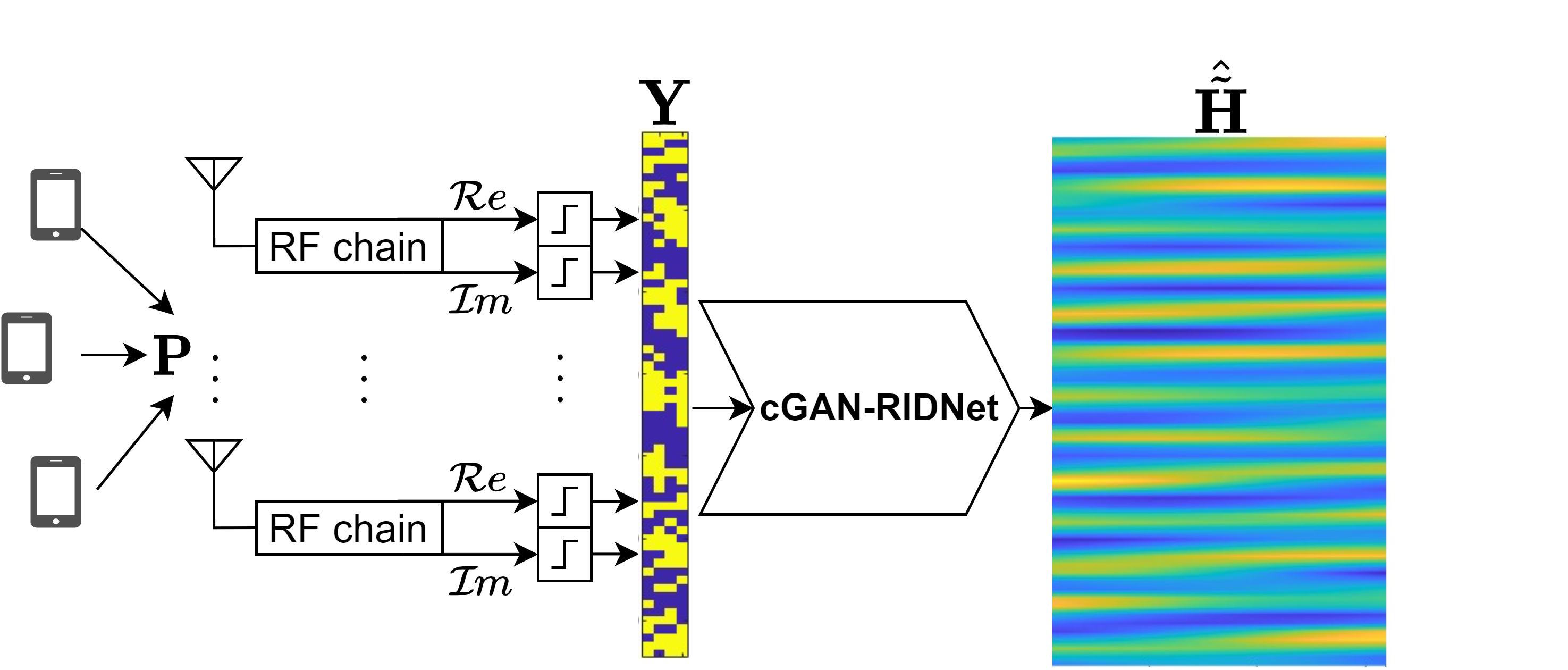}
    \caption{System model and channel estimation stages where $\mathbf{P}$ is pilot matrix, $\mathbf{Y}$ is collected signal and $\mathbf{\hat{\tilde{H}}}$ is estimated channel.}
    \label{fig:cgan_ridnet}
\end{figure}

We considered a mmWave massive MIMO system, where the \ac{bs} has $N$ \ac{rf} chains and $N$ antennas to serve $K$ single-antenna \ac{ue} simultaneously as depicted in Fig. \ref{fig:cgan_ridnet}. 
Also, two one-bit ADCs are connected to each RF chain.
The Saleh-Valenzuela channel model is employed to formulate the problem of one-bit \ac{adc} massive \ac{mimo} channel estimation.
The spatial channel vector $h_k$ with size $N \times 1 $ between the single \ac{ue} and $N$-antenna \ac{bs} can be expressed as 
\begin{equation}
\mathbf{h}_k=\sqrt{\frac{N}{L_k}} \sum_{l=1}^{L_k} \alpha^{(k,l)} \mathbf{a}\left(\theta_{a z}^{(k,l)}, \phi_{e l}^{(k,l)}\right),
\label{eq:spatial_channel}
\end{equation}
where $L_k$ is the number of channel paths while $\alpha^{(k,l)}$,  $\theta_{a z}^{(k,l)}$, and $\phi_{e l}^{(k,l)}$ indicate a complex gain, azimuth and elevation angles of $l^{th}$ path, respectively.
$\mathbf{a}\left(\theta_{a z}^{(k,l)}, \phi_{e l}^{(k,l)}\right)$ denotes an antenna array response matrix which varies by array geometry.
In this study, \ac{ula} is considered as follows
\begin{equation}
\mathbf{a}_{\mathrm{ULA}}(\theta_{az})=\frac{1}{\sqrt{N}}\left[e^{-j 2 \pi d \sin (\theta_{az}) \mathbf{n} / \lambda}\right],
\end{equation}
where $\mathbf{n}=[0,1, \cdots, N-1]^{T}$, 
$\lambda$ denotes wavelength, and $d = \lambda/2$  is antenna spacing. The spatial angles for the \ac{ula} can be defined as $\psi \triangleq d \sin (\theta_{az}) / \lambda$.
Finally the spatial channel matrix $\mathbf{H} \in \mathbb{C}^{N \times K}$ for all \ac{ue}s can be defined as 

\begin{equation}
\mathbf{H}=\left[\mathbf{h}_1, \mathbf{h}_2, \cdots, \mathbf{h}_k, \cdots, \mathbf{h}_K\right].
\end{equation}
A spatial channel can be transformed into an angular domain by utilizing a \ac{dft} matrix, which can be defined as
\begin{equation}
\mathbf{U}=\left[\overline{\mathbf{a}}_{\mathrm{ULA}}\left(\bar{\psi}^{1}_{az}\right), \overline{\mathbf{a}}_{\mathrm{ULA}}\left(\bar{\psi}^{2}_{az}\right), \cdots, \overline{\mathbf{a}}_{\mathrm{ULA}}\left(\bar{\psi}^{N}_{az}\right)\right]^H,
\end{equation} 
where $\bar{\psi}^{n}_{az}=\frac{1}{N}\left(n-\frac{N+1}{2}\right)$ for $n=1,2, \cdots, N$  is the grids for the angular domain which is predefined.
Then, the \ac{mm} channel in the angular domain can be defined as
\begin{equation}
    \mathbf{H^*} = \mathbf{UH}.
\end{equation}
RIDNet can exploit the sparsity of the mmWave channel by learning sparse features better.

\begin{figure*}[t!]
    \centering
    \includegraphics[width=1\linewidth]{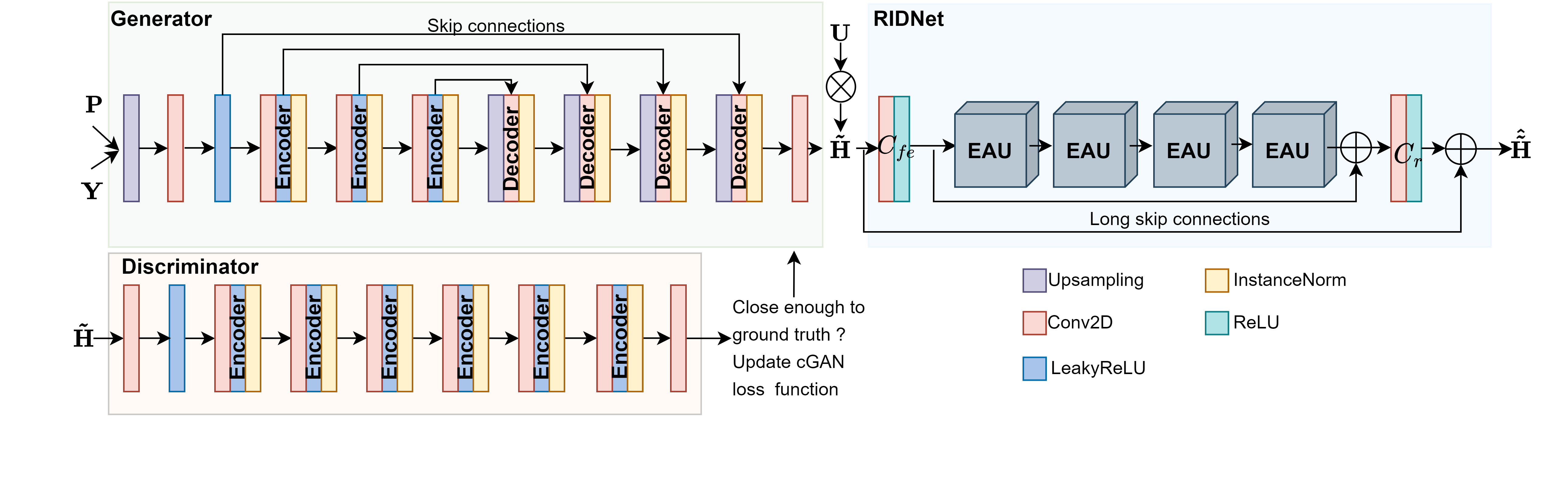}
    \caption{Network architecture of the suggested cGAN-RIDNet method.}
    \label{fig:overall_network}
\end{figure*}

The \ac{ue}s transmit a number of $Q$ pilots to the \ac{bs} during the training stage, and the received signal matrix $\mathbf{Y} \in \mathbb{C}^{N \times Q} $ using one-bit \ac{adc}s can be defined as
\begin{equation}
\mathbf{Y}=\operatorname{sgn}(\mathbf{H} \mathbf{P} +\mathbf{N})
\label{eq:received_signal},
\end{equation}
where $\mathbf{P} \in \mathbb{C}^{K \times Q}$ is pilot matrix, $\mathbf{N} \in \mathbb{C}^{N \times Q} $ denotes noise matrix whose elements follow $\mathcal{CN}\left(0, \sigma_n^2 \right)$, and $\operatorname{sgn(\cdot)}$ is an element-wise signum function utilized for both real and imaginary components of its input $x$ and defined as
\begin{equation}
\operatorname{sgn}(x)= \begin{cases}1 & , x \geq 0 \\ -1 & , \text { otherwise }.
\end{cases}
\end{equation}
An element of $\mathbf{Y}$ can take any value from the given array $\{1+j, 1-j,-1+j,-1-j\}$.

\section{ One-Bit \ac{adc} MIMO Channel Estimation Architecture}
In this section, the cGAN-RIDNet based one-bit ADC massive MIMO channel estimation method is introduced, which consists of two stages, namely c\ac{gan} and feature attention-based blind denoising network (RIDNet).  
In the first stage, the coarse \ac{mm} channel $\boldsymbol{\mathbf{\tilde{H}}}$ is estimated from the received signal $\mathbf{Y}$ by cGAN. The RIDNet in the second stage further enhances the channel estimation by reducing the noise effect over $\boldsymbol{\mathbf{\tilde{H}}}$ to obtain $\boldsymbol{\mathbf{\hat{\tilde{H}}}}$. 
The cGAN-RIDNet system architecture is shown in Fig. \ref{fig:overall_network}.  

\subsection{c\ac{gan}}
A conventional \ac{gan} architecture consists of two competitive models: generator and discriminator. 
The generator task discovers a random mapping from noise to actual data while the discriminator evaluates if the generated data is real or not; by doing so, two models are trained adversarially.
On the other hand, cGAN maps input samples to real data under given input conditions instead of random mapping.
Our cGAN model is adopted from \cite{isola2017image} with a modification in its loss function, where it is trained to discover the mapping relation between received signals $\mathbf{Y}$ and a channel matrix $\mathbf{H}$, conditioned on pilot symbols $\mathbf{P}$.

\subsubsection{Generator}
The U-Net structure, which enables pixel-level features to be preserved at various resolutions by utilizing skip connections, is employed in the generator. 
To reconstruct the channel from the reduced observations, firstly, the input is scaled by an upsampling layer. Therefore, three encoder and four decoder blocks are utilized in U-Net.
Each encoder and decoder block consists of convolutional and instance normalization layers. 
Convolutional layers of the decoders are used for upsampling of a given input to reconstruct the channel matrix, and instance normalization simplifies the learning process in generative models.
One important thing is that the generator can over-optimize for a certain discriminator feature, and the discriminator can not improve. 
As a result, the generator only produces outputs with a small variation to convince the discriminator. 
This is referred to as a mode collapse, and a $tanh$ activation function is used to prevent this phenomenon.

\subsubsection{Discriminator}
The patch structure is used in the discriminator, and it is composed of only convolutional layers. 
The patch structure is similar to a fully convolutional network.
Different from the traditional discriminators, rather than generating one label as real or fake, patch structure allows analyzing the input as different patches and evaluates from these patches to determine if the generator output is real or fake.
The patch structure consists of one convolutional layer with a $LeakyReLU$ activation function at the beginning, and it is followed by four encoder blocks.

\subsubsection{Loss Function}
The primary goal of the cGAN is for the generator to learn how to generate the most natural channel matrix conditioned on the pilot symbol matrix. At the same time, the discriminator gains the ability to differentiate between produced and real channel matrices.
Intuitively, the generator attempts to trick the discriminator by producing samples that appear indistinguishable from one another by optimizing the given loss function as
\begin{equation}
\begin{aligned}
\mathcal{L}_{\mathrm{cGAN}}(G , D, \mathbf{Y}, \mathbf{H}, \mathbf{P}) & =\mathbb{E}[\log D(\mathbf{H}, \mathbf{P})] \\
& +\mathbb{E}[\log (1-D(G(\mathbf{Y}, \mathbf{P}))].
\end{aligned}
\end{equation}
$G$ represents generator which aims to construct coarse channel matrices as $\mathbf{\tilde{{H}}}$ and $D$ denotes discriminator which aims to identify the generated channel $\mathbf{\tilde{H}}$ from the actual channel $\mathbf{{H}}$.
Thus, it is a min-max game between the generator and the discriminator, denoted as
\begin{equation}
\min _\Psi \max _\Theta \mathcal{L}_{\mathrm{cGAN}}\left(G_\Psi, D_\Theta, \mathbf{Y}, \mathbf{H}, \mathbf{P}\right).
\end{equation}
The generator aims to minimize this function, while the discriminator tries to maximize it by optimizing corresponding parameters $\Psi$ and $\Theta$, respectively.

The $\mathcal{L}_1$ and $\mathcal{L}_2$ regularizations are added to cGAN's loss function to generate a more realistic channel matrix, and they are expressed as
\begin{equation}
\mathcal{L}_1=\mathbb{E}\left[\left\|\mathbf{H}-G(\mathbf{Y}, \boldsymbol{P})\right\|\right],
\end{equation}
and
\begin{equation}
\mathcal{L}_2=\mathbb{E}\left[\left\|\mathbf{H}-G(\mathbf{Y}, \boldsymbol{P})\right\|^2\right],
\end{equation}
respectively.
$\mathcal{L}_1$ regularization allows to retain of pixel-level details and makes the generated samples less blurry.
However, it can cause sharp transitions between channel matrix elements.
With $\mathcal{L}_2$ regularization, the discriminator's task remains the same; however, the generator is tasked to generate smoother samples. Thus, a good balance can be found to create a more realistic channel. 
The overall loss function becomes 
\begin{equation}
\min _\Psi \max _\Theta \mathcal{L}_{cG A N}\left(G_\Psi, D_\Theta, \mathbf{Y}, \mathbf{H}, \mathbf{P}\right)+\lambda_1\mathcal{L}_1+\lambda_2\mathcal{L}_2,
\end{equation}
with $\lambda_1$ and $\lambda_2$ are weighting coefficients for $\mathcal{L}_1$ and $\mathcal{L}_2$ regularization, respectively

 Following coarse reconstruction of the one-bit \ac{adc} \ac{mimo} channel matrix $\mathbf{\tilde{H}}$ from $\mathbf{Y}$ and $\mathbf{P}$ with cGAN, the estimated channel is transferred to the DL-based denoising network.
\subsection{Feature Attention Based Blind Denoising Network}
To enhance channel estimation performance, we utilized a feature attention-based blind denoising network. 
The formerly reconstructed channel with the cGAN is forwarded to the RIDNet whose architecture~\cite{anwar2019real} consists of three key modules: the feature extraction module ($C_{fe}$), the feature learning residual on the residual module ($R_{fl}$), and the reconstruction module ($C_r$), respectively. 
The initial features from noisy channel matrix $\mathbf{\tilde{H}}$ can be extracted by using a feature extraction module, which consists of one convolutional layer and is represented as

\begin{equation}
f_{H_0}=C_{fe}({\mathbf{\tilde{H}}}),
\end{equation}
where $C_{fe}(\cdot)$ refers to the process of convolution performed on the input, and the resulting acquired features are denoted by $f_{h_0}$.
Next, the extracted features are passed on to $R_{fl}$, which comprises four successive \ac{eau} units, each consisting of residual blocks with short and local skip connections. We will explain the role of this module later. By incorporating $R_{fl}$, the network can better learn the underlying patterns in the input data, which enables it to identify sparse features and estimate the noise in the channel matrix.
This module can be represented as
\begin{equation}
f_{H_r}=R_{f l}\left(f_{H_0}\right),
\end{equation}
where  $f_{H_r}$ denotes the explored features. 
Then $f_{H_r}$ is summed up with the output of the first convolutional layer by long skip connection and forwarded to the reconstruction module, which consists only a convolutional layer $C_r(\cdot)$
\begin{equation}
\mathbf{\tilde{R}}=C_{r}\left(f_{H_r}\right).
\end{equation}
To capture residual noise, the model uses a long skip connection that connects the network input to the output, as opposed to only mapping input data to a denoised output.
\begin{equation}
\mathbf{\hat{\tilde{{H}}}}= \mathbf{\tilde{H}} + \mathbf{\tilde{R}},
\end{equation}
where $\mathbf{\hat{\tilde{{H}}}}$ is the output of the whole network and represents the estimated channel. 

The baseline characteristics can be managed to keep for the regression because of the utilized two long-skip connections in the network, which allow initially obtained information to be conveyed directly to the end of the network.
The loss function of the network for the given training set $\{\tilde{\mathbf{H}}_i, \mathbf{H}_i\}_{i=1}^M$, where the input for the RIDNet is the result of the \ac{cgan} and the label is the \ac{mm} \ac{mimo} channel, can be expressed as
\begin{equation}
L(\boldsymbol{\mathcal{W}})=\frac{1}{M} \sum_{i=1}^{M}\left\|\mathcal{F}_{\operatorname{RIDNet}}\left(\tilde{\mathbf{H}}_i \right)-\mathbf{{H}}_{i}\right\|_{1},
\end{equation}
where $\boldsymbol{\mathcal{W}}$ denotes all the trainable parameters of the network and $\mathcal{F}$ denotes mapping function.

As a result, we can represent the final outcome of the proposed cGAN-RIDNet as
\begin{equation}
    \mathbf{\hat{\tilde{{H}}}}=\mathcal{F}_{\operatorname{RIDNet}}\left( \mathcal{F}_{\operatorname{cGAN}}( \boldsymbol{{\tilde{{\mathbf{H}}}}} ;\mathbf{Y}, \mathbf{P}) \right).
\end{equation}

\subsubsection{Enhanced Attention Unit}
Two convolutional layers in the EAU are used to expand the pre-estimated features into two branches.
The output of the branches is then concatenated and transmitted via a convolutional layer, which is followed by two consecutive residual blocks ($RB$).
The first $RB$ has two convolutional layers and allows the extraction of features, but the second $RB$ has three convolutional layers and allows comprehension by flattening with an additional convolutional layer.
It is worth noting that convolutional layers often allow for the extraction of local characteristics.
After the residual blocks in the EAU module, a global average pooling layer is employed to retain the global features.
This allows for the exploration of sparse channel features as well as inconsistencies in adjacent elements of the pre-estimated channel matrix.
The global features are fed into two convolutional layers where the sigmoid activation function is used at the edge.
The channel features captured at the end of the second residual block are multiplied with the result of the sigmoid activation function to utilize the statistics acquired through global average pooling, as described in \cite{anwar2019real}.
By employing a short skip connection, the product of the multiplication is combined with EAU's input, enabling captured features and characteristics to flow between units.
\begin{figure}
    \centering
    \includegraphics[width=0.9\linewidth]{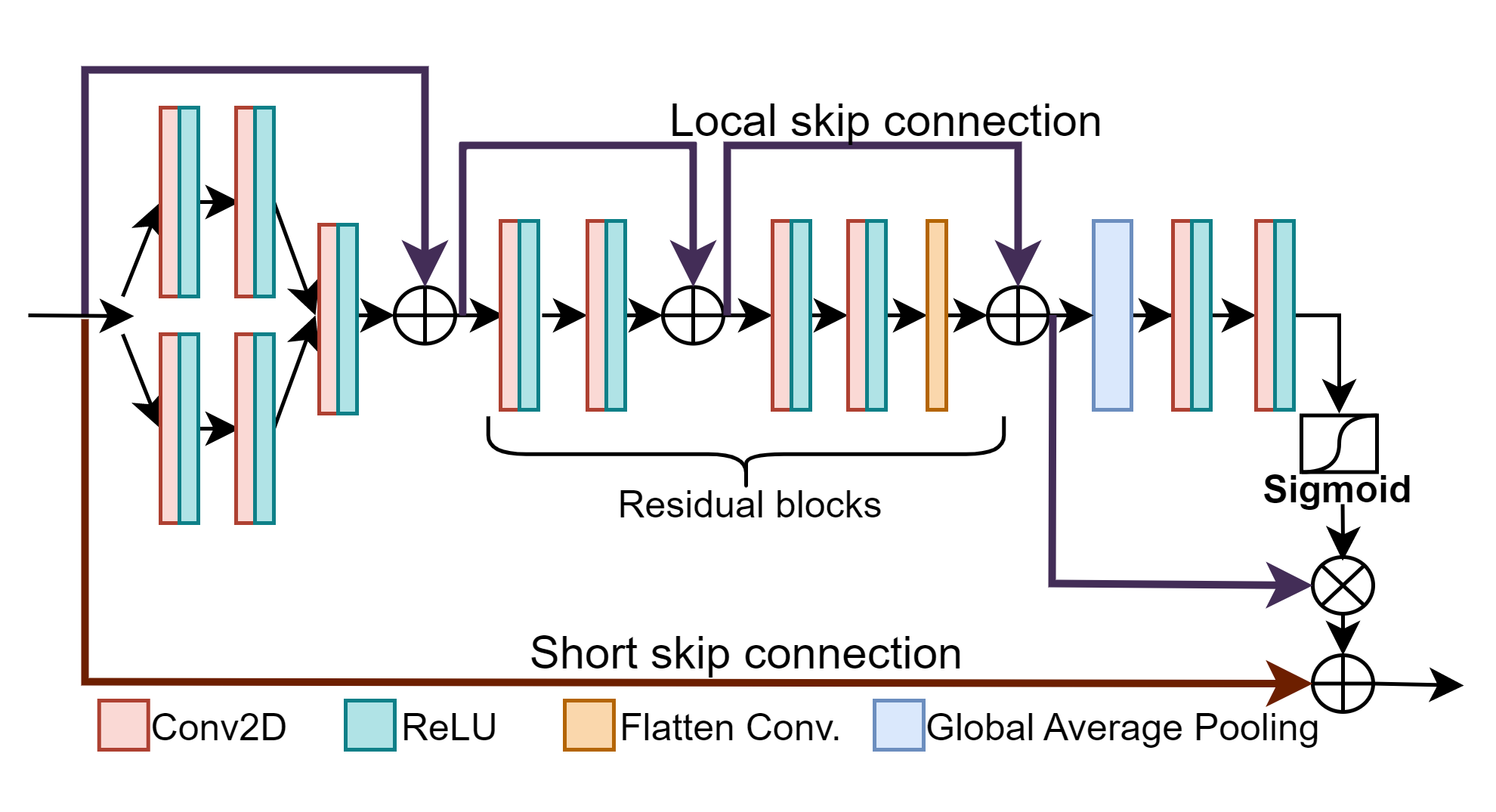}
    \caption{
    The structure of the enhanced attention unit involves the utilization of two residual blocks.\cite{anwar2019real}.}
    \label{fig:eau}
\end{figure}

\section{Simulations and Results}
This section outlines the initialization settings and training strategies used for cGAN and RIDNet, followed by the presentation of simulation results showcasing the effectiveness of the proposed approach.
We assume $K=32$, and each \ac{ue} has an antenna. \ac{bs} assumed to have $N={64,128,192,256}$ antennas, which are connected to two one-bit ADCs to sample real and imaginary components of the received signal.
Each \ac{ue} channel is assumed to be composed of up to ten paths ($L_k = 10$).  
We set the remaining parameters of the channel in Eq. \eqref{eq:spatial_channel} for each UE as follows; for $l^{th}$ path as  $\alpha^{(k,l)} \sim \mathcal{CN}(0,1)$, and $\theta_{a z}^{(k,l)} \sim\mathcal{U}\left(-\frac{1}{2}\pi, \frac{1}{2}\pi\right)$.
\subsection{cGAN-RIDNet Initialization}

Each convolutional layer in the generator has $128$ filters with $4\times4$ kernel size.
Convolution layers in the discriminator contain 512 filters with $4 \times 4$ kernel size.
Only the last layer of the discriminator contains a single filter to get an overall result from the patches.

Convolutional layers of the RIDNet have $64$ filters and are unified with the $ReLU$ activation function.
The convolutional layers in the model have kernel sizes of $3\times3$, except for the layer used to flatten inside the EAU, which has a kernel size of $1 \times 1$. Four cascaded EAU modules were used for improved feature extraction.
\subsection{Training}
$30 \times 10^4$ data pairs were used to train both networks for SNR values of $[-10,20]$ by 5 dB margin, and for validation, $\%20$ of the training data was reserved.  
The training data consist of two channels where the channels contain real and imaginary samples.
cGAN model is trained using the MSProp algorithm with  $2 \times 10^{-4}$ and $2 \times 10^{-5}$ learning rates for the generator and discriminator, respectively.
In training RIDNet, the Adam optimizer is employed with an initial learning rate of $10^{-4}$, which is later decreased to $10^{-5}$ during the training process, and eventually set to $10^{-6}$ to faster convergence.
The TensorFlow framework is used to construct the network architecture and train the network offline. 
To investigate the effect of the sparse depiction of the \ac{mm} channel, we also converted the initially estimated coarse channel matrix $\mathbf{\tilde{H}}$ to angle domain $\mathbf{\tilde{H}}^*=\mathbf{U}\mathbf{\tilde{H}}$ with DFT matrix $\mathbf{U}$. Then, the RIDNet also trained with transformed pairs $\{{\mathbf{\tilde{H}}_i^*}, \mathbf{H}_i^*\}_{i=1}^{M}$.

\subsection{Comparison}
The performance of the proposed method is assessed by employing the normalized mean square error (NMSE), as defined below
\begin{equation}
    \mathrm{NMSE (dB)}=10 \times log_{10} \left( \mathbb{E}\left\{
    \frac{\|\hat{\tilde{\mathbf{H}}}-{\mathbf{H}}\|_{2}^{2}}{\|{\mathbf{H}}\|_{2}^{2}}\right\} \right).
\end{equation}

\begin{figure}[ht!]
    \centering
    \includegraphics[width=1\linewidth]{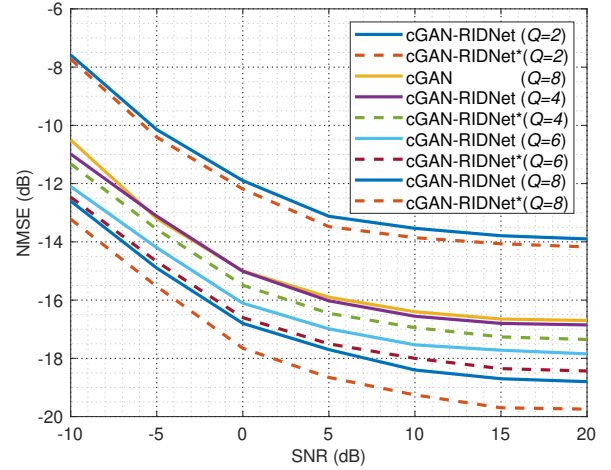}
    \caption{The channel estimation performance comparison of the proposed method with different pilot transmissions.}
    \label{fig:estimation_performance_pilot}
\end{figure}

We evaluated the NMSE of the proposed cGAN-RIDNet to that of cGAN for various pilot transmission instants (i.e., $Q={2,4,6,8}$), and the results are presented in Fig. \ref{fig:estimation_performance_pilot}.
Compared to cGAN, the employment of cGAN-RIDNet leads to a substantial enhancement in channel estimation performance across all SNR regimes for the same number of pilot transmissions $Q=8$.
With the suggested approach, it is possible to achieve the cGAN's channel estimation performance at $15$ dB SNR at about $1$ dB SNR.
Also, even when just half of the pilot transmission is used ($Q = 4$), the cGAN-RIDNet channel estimation performance almost exceeds the cGAN ($Q = 8$) performance.
Furthermore, we investigated the effect of the sparse representation of the \ac{mm} channel. 
It is analyzed that the RIDNet becomes more reliant and capable of disclosing features in the channel more effectively with this transformation because the RIDNet design can focus on sparse characteristics owing to its attention mechanism.
Consequently, a straightforward transformation offers the suggested technique an additional channel estimation performance gain for all pilot transmissions. 
When there is only $Q=2$ pilot transmission, the channel estimation performance of the proposed method degrades significantly.
This is due to insufficient data to identify channel properties accurately.

\begin{figure}[ht!]
    \centering
    \includegraphics[width=1\linewidth]{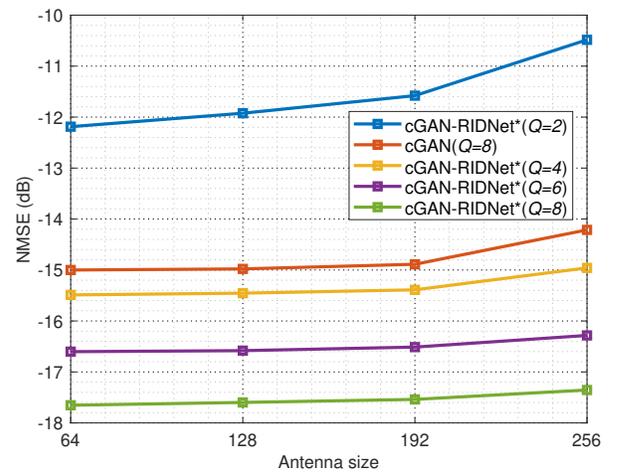}
    \caption{Channel estimation performance for different antenna sizes.}
    \label{fig:estimation_performance_antenna}
\end{figure}
\vspace{10pt}
We also evaluated the performance of the proposed method for the cases where the number of pilots is constant, but the number of antennas increases, as shown in Fig. \ref{fig:estimation_performance_antenna}. The channel estimation performance of the proposed method significantly improves for all antenna sizes over the cGAN, even when only half of the pilot transmission is used. 
Even with a relatively modest pilot size ($Q=4,6,8$), the performance of the cGAN-RIDNet remains reliable even when more antennas are installed at the BS.

\begin{figure}[ht!]
    \centering
    \includegraphics[width=1\linewidth]{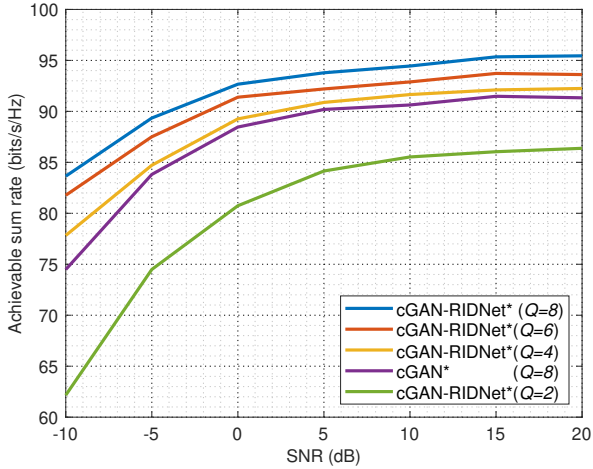}
    \caption{The achievable sum rate for $K=32$ \ac{ue} with interference-aware beam selection scheme~\cite{gao2016near} under the proposed channel estimation method.}
    \label{fig:asr}
\end{figure}

Furthermore, Fig. \ref{fig:asr} demonstrates the achievable sum rate performance with the interference-aware beam selection scheme.
As predicted by the previous findings, the proposed approach can achieve higher throughput using a lower pilot transmission overhead. 

\begin{table}[ht!]
\renewcommand{\arraystretch}{1.4}
\setlength{\tabcolsep}{6pt}
\centering
\caption{Computation times of the networks}
\label{tab:processing_times}
\begin{tabular}{|c|c|c|}
\hline
\textbf{cGAN} & \textbf{RIDNET} & \textbf{cGAN-RIDNET} \\ \hline
26.023 ms     & 0.2267 ms       & 26.2497 ms           \\ \hline
\end{tabular}
\vspace{10pt}
\end{table}

Finally, the computation times of the cGAN and RIDNET networks separately and the proposed cGAN-RIDNet approach are presented in Table \ref{tab:processing_times}. 
Since the RIDNet requires a significantly low computational load compared to the cGAN, the overall computation time of the cGAN-RIDNet is quite close to the cGAN.
Considering the accuracy improvement for the channel estimation performance by the cGAN-RIDNET, this minor increase in the computation time is a reasonable compromise, and the proposed method can be used in practical scenarios.

\section{Conclusion}
Our focus was on investigating the issue of estimating the \ac{mm} \ac{mimo} channel in situations where the \ac{bs} employs one-bit \ac{adc}s.
The channel estimation performance of the one-bit \ac{mm} \ac{mimo} can be improved by employing cGAN-RIDNET, which is leveraged by both cGAN and the feature attention-based blind denoising network. 
The numerical results demonstrate that the proposed method significantly enhances the channel estimation accuracy using a lower pilot overhead while computation time is kept comparable.

\section*{Acknowledgment}
This publication was made possible in parts by  NPRP13S-0130-200200  and by NPRP14C-0909-210008 from the Qatar National Research Fund (a member of The Qatar Foundation). The statements made herein are solely the responsibility of the authors.
We thank to StorAIge project that has received funding from the KDT Joint Undertaking (JU) under Grant Agreement No. 101007321. The JU receives support from the European Union’s Horizon 2020 research and innovation programme in France, Belgium, Czech Republic, Germany, Italy, Sweden, Switzerland, Türkiye, and National Authority TÜBİTAK with project ID 121N350.
\balance
\bibliographystyle{IEEEtran}
\bibliography{main.bib}

\vspace{12pt}
\color{red}

\end{document}